\documentclass{elsart41}
\usepackage{graphics}
\usepackage{graphicx}
\usepackage{amssymb}
\begin{document}

\begin{frontmatter}



\title{Thermoelectric effects in strongly interacting quantum dot coupled to 
       ferromagnetic leads}
%---- Don't remove this comment line! ----

\author{Mariusz Krawiec},
\ead{krawiec@kft.umcs.lublin.pl}
\author{Karol I. Wysoki\'nski\corauthref{Karol}},\,
\ead{karol@tytan.umcs.lublin.pl}
%\author[BB]{C. Name3}
\address{Institute of Physics and Nanotechnology Center, M. 
         Curie-Sk\l{}odowska University, Pl. M. Curie-Sk\l{}odowskiej 1,
         PL~20-031~Lublin, Poland}  
\corauth[Karol]{Corresponding author. Tel: +48815376236, fax:+4881 5376191}

\begin{abstract}
We study thermoelectric effects in Kondo correlated quantum dot coupled to 
ferromagnetic electrodes by calculating thermopower S in the Kondo regime as 
function of on-dot energy level and temperature. The system is represented by 
the Anderson model and the results agree well with those recently measured for 
a quantum dot coupled to nonmagnetic leads. For magnetic electrodes one 
observes marked dependence of  S on the degree of their polarization.  
\end{abstract}
\begin{keyword}
quantum dot \sep thermopower \sep Kondo effect
% PACS codes here, in the form: 
\PACS    75.20.Hr, 72.15.Qm, 72.25.-b, 73.23.Hk 
\end{keyword}
\end{frontmatter}
% main text

Transport properties of artificial nanostructures with highly correlated 
electrons present fruitful research area. In particular the many body effects 
have been predicted \cite{Glazman} and later observed \cite{Goldhaber-Gordon} 
in quantum dots coupled to external leads. Such structures allow study of 
various physical phenomena in well controlled conditions and geometries. Number 
of possibilities have already been tested experimentally resulting in 
discoveries of new effects \cite{PW}.

In this contribution we are interested in the study of the voltage $V_T$ across 
quantum dot induced by the temperature gradient characterised by $\Delta T$. 
The ratio $V_T \over \Delta T$ is a measure of the thermopower $S$. We consider single quantum dot with only one energy level 
coupled to two external ferromagnetic leads. The second electron on the dot 
experiences strong repulsion $U$ due to charging energy $e^2/2C$, where $C$ is 
the capacitance of the dot. We model the system by the $U = \infty$ single 
impurity Anderson Hamiltonian in the slave boson representation and adopt the 
calculation scheme and the approximations presented previously in \cite{MK}. 
\begin{eqnarray}
H = \sum_{\lambda {\bf k} \sigma} \epsilon_{\lambda {\bf k} \sigma} 
c^+_{\lambda {\bf k} \sigma} c_{\lambda {\bf k} \sigma} 
+ \sum_{\sigma} E_{d} f^+_{\sigma} f_{\sigma} \nonumber \\
+ \sum_{\lambda {\bf k} \sigma} \left( V_{\lambda {\bf k} \sigma} 
c^+_{\lambda {\bf k} \sigma} b^+ f_{\sigma} + H. c. \right),
\label{Hamilt}
\end{eqnarray}
where $\lambda = L$ ($R$) denotes left (right) lead, 
$c^+_{\lambda {\bf k} \sigma}$ ($c_{\lambda {\bf k} \sigma}$) 
is the creation (annihilation) operator for a conduction electron with 
the wave vector ${\bf k}$, spin $\sigma$ in the lead $\lambda$ 
and $V_{\lambda {\bf k} \sigma}$ is the hybridization matrix element between 
localized electron on the dot with the energy $E_d$ and conduction electron of 
energy $\epsilon_{\lambda {\bf k}}$ in the lead $\lambda$. 

The non-equilibrium Green's function technique \cite{Keldysh} allows one to 
study the general situation with arbitrary value of the $\Delta T$. Here, 
however, we limit our consideration to the linear response. This means that the 
response coefficients $L_{ij}$ between charge and heat fluxes and chemical 
potential and temperature gradients are evaluated for an equilibrium 
($\mu_L=\mu_R$) system with constant temperature ($T_L=T_R$). Electrical 
conductance $G$ is given by the coefficient $L_{11}$ as 
$G={-e^2\over T}L_{11}$, while thermopower requires knowledge of both  $L_{11}$ 
and $L_{12}$ and is given by $S={-1\over {eT}}{L_{12}\over L_{11}}$. The linear 
response coefficients read 
\begin{eqnarray}
L_{11} = \frac{T}{h} \sum_{\sigma} \int d\omega \Gamma_{\sigma} (\omega) 
{\rm Im} G^r_{\sigma}(\omega) 
\left(-\frac{\partial f(\omega)}{\partial \omega} \right)_T,
\label{L11}
\end{eqnarray}
\begin{eqnarray}
L_{12} = \frac{T^2}{h} \sum_{\sigma} \int d\omega \Gamma_{\sigma} (\omega) 
{\rm Im} G^r_{\sigma}(\omega) 
\left(\frac{\partial f(\omega)}{\partial T} \right)_{\mu}.
\label{L12}
\end{eqnarray}
\begin{figure}[ht]
\begin{center}
\includegraphics[width=0.38\textwidth]{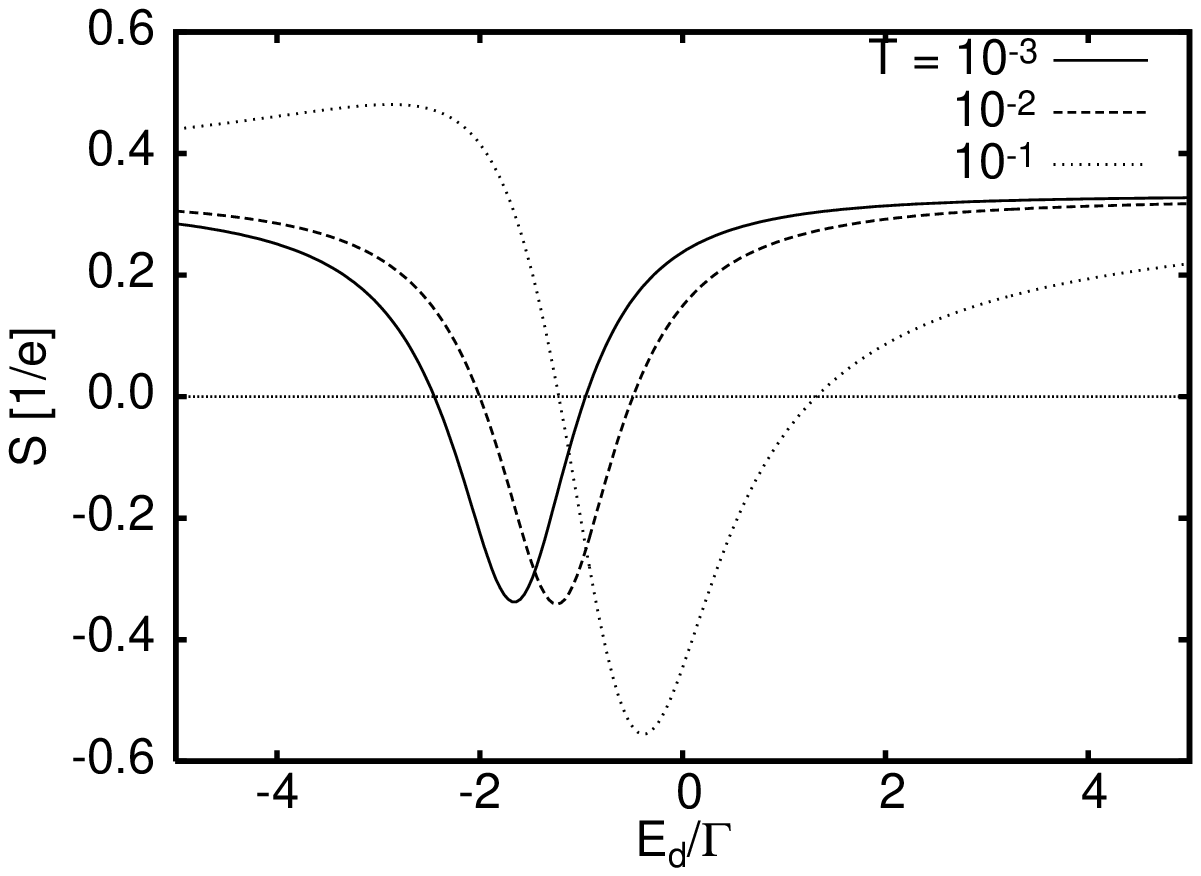}
\includegraphics[width=0.38\textwidth]{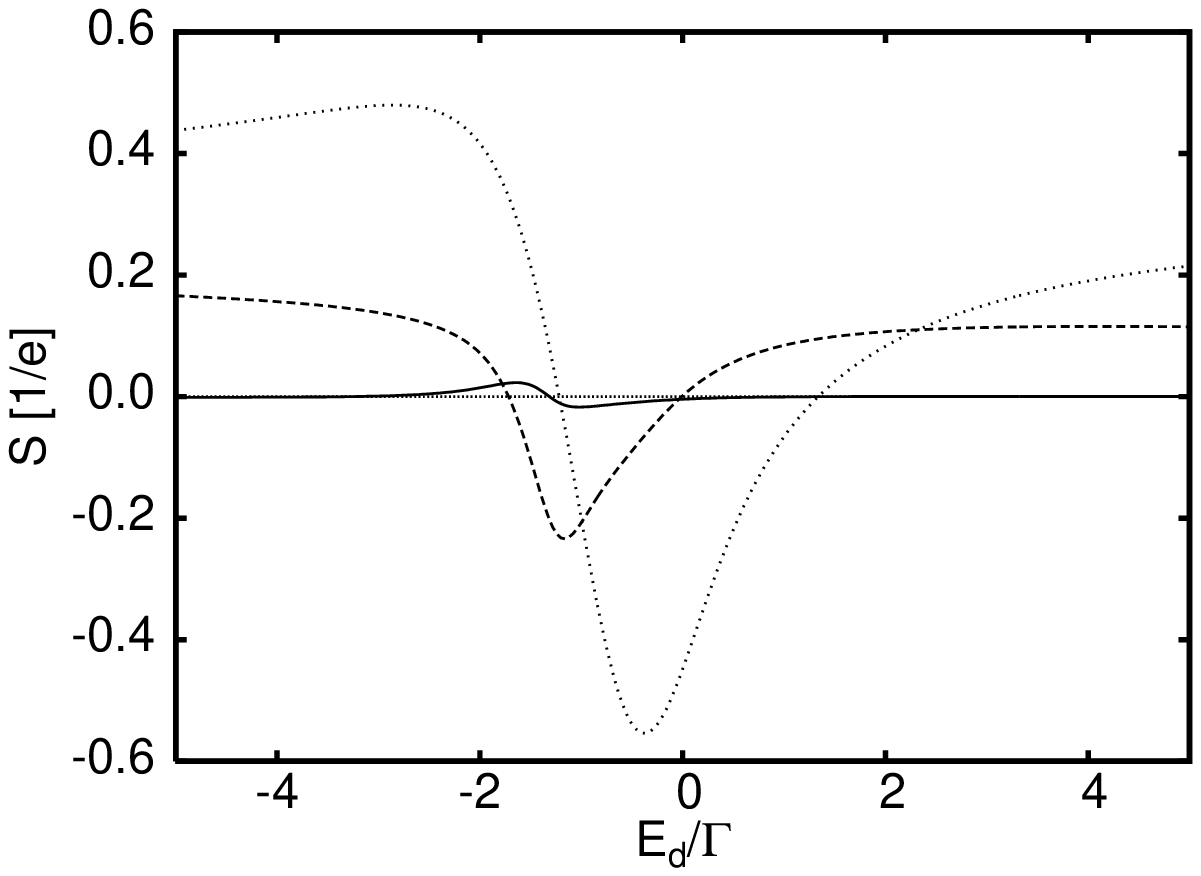}
\end{center}
\caption{The linear response thermopower of the quantum dot coupled to 
         magnetically polarised leads as a function of the position of the 
	 on-dot electron energy calculated for different temperatures (in units 
	 of $\Gamma$): $T=10^{-3}$, $T=10^{-2}$  and $T=10^{-1}$. Polarisation 
	 of the leads $P=0.0$ (upper panel) and $P=0.1$ (lower panel). 
	 Splitting of the Kondo resonance results in a very small thermopower.} 
\label{fig1}
\end{figure}
\begin{figure}[ht]
\begin{center}
\includegraphics[width=0.38\textwidth]{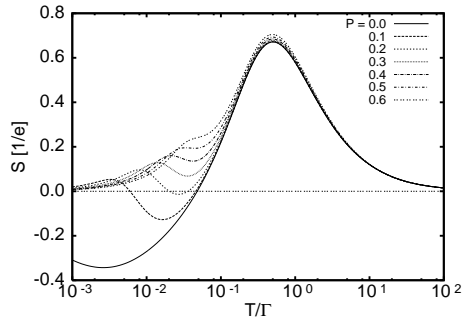}
\end{center}
\caption{The temperature dependence of the thermopower of the quantum dot 
         coupled to magnetically polarised leads with P=0.0, 0.1, 0.2, 0.3, 
	 0.4, 0.5, 0.6 (starting from lowest curve near the deep minimum).}
\label{fig2}
\end{figure}

As it has already been noted in \cite{Boese} in the context of N-QD-N system, 
the thermopower S is a powerful characteristic to identify the appearance of 
the Kondo effect as it changes sign when the system is cooled  down to 
temperatures $T$ below $T_K$. This is particularly clear in Fig. \ref{fig1} and 
in Fig. \ref{fig2} (solid curve), where S vanishes for temperature equal actual 
Kondo temperature $T_K$. The behaviour is more complicated in the  FM-QD-FM 
system with nonzero polarisation (lower panel of Fig. \ref{fig1} and Fig.
\ref{fig2}). 
 
It is worth to note that the thermopower of the quantum dot defined in the two 
dimensional electron gas has recently been measured experimentally 
\cite{Scheibner} and qualitatively agrees with that presented in figure
\ref{fig1} for $P=0$. To understand their results the authors \cite{Scheibner} 
use the Mott formula, which states that thermopower is proportional to the 
logarithmic derivative of the conductance with respect to the energy evaluated 
at the actual Fermi energy. The Kondo effect shows up as additional maximum 
appearing in conductance $via$ quantum dot when the temperature decreases. It 
is connected with the Abrikosov-Suhl resonance (resonances) in the density of 
states, which appears (appear) at low temperatures and is (are) located at the 
position(s) of the Fermi level of the leads in (non)equilibrium situation. The 
slope of conductance thus changes and the thermopower changes sign. Thermopower 
measured directly and obtained from the Mott formula differ in line shapes 
\cite{Scheibner} and this calls for the discussion of the validity of Mott 
formula \cite{Lunde} in the interacting quantum dot systems. This as well as 
detailed comparison between experimental data and calculations will be the 
subject of future work. 
 
In magnetically polarised systems the Abrikosov - Suhl resonance splits (as in 
external B field \cite{Goldhaber-Gordon}) and this results in decrease of low 
temperature S (Fig. \ref{fig1}). 

\vspace{-2mm}
\section*{Acknowledgment}
This work has been supported by the Polish MNiI under the contract 
PBZ-MIN-P03/2003.

\end{document}